\documentstyle[preprint,revtex,eqsecnum]{aps}
\begin{document}
\tolerance 50000
\preprint{
\begin{minipage}[t]{1.8in}
UCSBTH-93/37\\[-0.8em]
LPQTH-93/20
\end{minipage}
}
\draft
\vspace*{0.5cm}

\begin{center}
{\bf RESONANT IMPURITY SCATTERING IN A STRONGLY CORRELATED ELECTRON
MODEL}
\end{center}

\vspace*{0.45cm}
\author{D. POILBLANC\cite{byline1}, W. HANKE\cite{byline2},
and D. J. SCALAPINO}
\vspace*{0.2cm}
\begin{instit}
\begin{center}
Department of Physics,\\
University of California at Santa Barbara \\
Santa Barbara, CA 93106
\end{center}
\end{instit}
\vspace*{0.2cm}

\receipt{\hskip 4truecm}

\begin{abstract}

Scattering by a single impurity introduced in a strongly correlated
electronic system is studied by
exact diagonalization of small clusters.
It is shown that an inert site which is spinless and unable to accomodate holes
can give rise to strong resonant
scattering. A calculation of the local density of state reveals that,
for increasing antiferromagnetic
exchange coupling, d, s and p-wave symmetry bound states in which
a mobile hole is trapped
by the impurity potential induced by a local distortion of the
antiferromagnetic background
successively pull out from the continuum.

\end{abstract}
\vspace*{2cm}

\pacs{PACS numbers: 74.72.-h, 71.27.+a, 71.55.-i}

The nature of impurity scattering in the high-$T_c$ copper oxide
superconductors
is of particular interest for understanding their low-temperature transport
properties. For example, recent experimental studies \cite{Hardy} have found
that $Zn$ impurities can change the low temperature dependence of the
penetration depth from a linear to a quadratic temperature variation. A similar
behavior is seen in the temperature dependence of the Knight
shift \cite{Ishida}. In addition, the transport lifetime observed in microwave
surface resistance measurements is impurity limited at low temperatures
\cite{Bonn}. It is believed that $Zn$ goes into a planar Cu(2) site suppressing
the local moment on its site. Theoretical calculations of various transport
properties \cite{Hirschfeld1,Hotta,Hirschfeld2} have found that models
which assume a $d_{x^2-y^2}$ gap with strong resonant impurity scattering
provide a possible explanation for the experimental data. However, the origin
of
this resonant scattering remains an
open question. In particular, is such
resonant scattering by local defects a consequence of the strong correlations
in
the host system? Here we report the results of numerical calculations on a t--J
model with an inert impurity. We find that an added hole can have boundstates
of various symmetries as J/t increases. Thus as the host J/t ratio increases,
a local inert impurity can give rise to strong resonant scattering.

Now, as one knows, a local impurity introduced into a non-interacting electron
system can give rise to boundstates \cite{Callaway}. Thus a single on-site
repulsive (attractive) one-body impurity potential in a tight-binding lattice
leads to a boundstate located in energy above (below) the band when the
strength
of the potential exceeds a critical value. In the calculations
of transport properties, the effects of impurity scattering are often
characterized by scattering phase shifts. For energies near a boundstate
the phase shift approaches $\pi /2$ \cite{JFsumrule}
and one has strong resonant scattering. In this case the nature of the
scattering is directly related to the one-body impurity potential.
Here we are interested in the problem of an impurity in a strongly interacting
host.
The central idea we would like to put forward in this paper is the fact that
an impurity introduced in a strongly correlated ground state (GS) could behave
quite differently from an impurity in a weakly interacting system.
In the case of an on-site impurity potential in a tight binding lattice
mentioned above, the interaction
is spacially located at the perturbed site, and the resultant bound states
can have only s-wave symmetry. However, in a
many-body ground state like an antiferromagnet (AF), the antiferromagnetic
correlations
are slightly enhanced in the vicinity of a vacancy \cite{DJSimpurity}, and thus
the impurity produces a dynamic finite range potential.
Bound states of various symmetries
can then result as we shall show in this paper. Furthermore the occurance of
these states depends upon the strength of the correlations in the host.

An inert site introduced in a two-dimensional
AF background can be described by the following hamiltonian,
\begin{equation}
{\rm H =
{\rm J } \sum_{{\bf \langle i j\rangle }}^{\quad\quad\prime}
( {{\bf S}_{\bf i}}.{{\bf S}_{\bf j}} - {1\over4} n_{\bf i} n_{\bf j} )
- {\rm t} \sum_{{\bf \langle i j \rangle},s}^{\quad\quad\prime}
P_G({c}^{\dagger}_{{\bf i},s} {c}_{{\bf j},s} + h.c.)P_G },
\label{hamiltonian}
\end{equation}
\indent
where the notations are standard and the prime means that the sum
over the nearest neighbor links ${\bf \langle i j\rangle }$ is
restricted to the bonds {\it not} connected to the impurity site.
The kinetic term of (\ref{hamiltonian})
describes the motion of extra empty or doubly occupied sites
in the large-U limit of the Hubbard model.
For simplicity, in the following we shall call "holes" either of
these entities (${c}^{\dagger}_{{\bf i},s}$ is the hole {\it creation}
operator).
The impurity model can be obtained continuously from the uniform model
(large-U Hubbard or t--J models) by gradually
turning off the hoppings on the 4 bonds connected to a given site O.
Eventually for vanishing couplings the impurity spin
becomes frozen e.g. $S_O^z=-1/2$ (formally, this
can be achieved by adding an infinitesimal magnetic field). In other words, if
a
vacant site is introduced in a strongly correlated host by, let
us say, removing a down spin at site O then an excess $S^z=1/2$ is left over
as depicted schematically in Fig. 1.
In this case the scattering of an extra hole added to the system
can occur in two different spin channels $S^z=0$ or $S^z=1$. However, since the
$S^z=1$ sector involves triplet states of higher energies
we shall restrict ourselves later on mainly to the $S_z=0$ scattering channel.
The following results are obtained by exactly diagonalizing small
$4\times 4$, $\sqrt{18}\times \sqrt{18}$,  $\sqrt{20}\times \sqrt{20}$
and  $\sqrt{26}\times \sqrt{26}$ clusters by a standard Lanczos method.
We will measure energies in unit of t.

Let us now formally construct the impurity state by removing a spin $-\sigma_0$
at site O from the AF and having the spin system relax around the impurity.
The local hole density of states at a site $\bf i$
away from the impurity site is given by
\begin{equation}
N_{\bf ii}^{\sigma,\sigma_0}(\omega)=
-\frac{1}{\pi} Im\, \big< \Psi_{0,\sigma_0} ^{imp} \mid c_{\bf i,\sigma}
\frac{1}{\omega+i\epsilon-H+E_0^{imp}} c_{\bf i,\sigma}^\dagger \mid
\Psi_{0,\sigma_0}^{imp}\big> ,
\label{density}
\end{equation}
\indent
with $\mid\Psi_{0,\sigma_0}^{imp}\big>$ the impurity GS.
Note that the energies are measured with respect to the GS energy at
half-filling $E_0^{imp}$. An equivalent expression for
the local density of the pure AF system is obtained by
replacing "imp" by "pure" and omitting $\sigma_0$.
In the pure case, an interesting structure
appears at the bottom (top) of the upper (lower) Hubbard band with increasing
coupling J. Indeed, recent exact calculations on various cluster
sizes \cite{DP1holespec} have
shown that a quasi-particle band of width $\sim J$
survive with increasing system size (see also Fig. 2f). Our first
motivation here is to investigate the influence of the impurity on this
band structure.

At this stage it is useful to notice that the local density of states
obeys the following sum rule,
\begin{equation}
\int_{-\infty}^{+\infty} N_{\bf ii}^{\sigma,\sigma_0}(\omega)\,d\omega =
\frac{1}{2}+\sigma\big< S_{\bf i}^z\big>_0,
\label{sumrule}
\end{equation}
\indent
where $\big< ...\big>_0$ stands for the expectation value in the AF GS.
In the pure system, a small tunneling between the two N\'eel GS (for a
finite system) leads to a zero expectation value of $S_{\bf i}^z$ for any site.
However, the impurity breaks translation symmetry and the removal of
a spin at site O imposes a local AF spin environment around it.
In this sense the vacant site acts like a magnetic impurity of spin
$-\sigma_0$.
The local density of states
$N_{\bf ii}^{\sigma,\sigma}$ at all non-equivalent sites of a 20-site
cluster at an intermediate coupling J=0.5
is shown on Figs. 2a--e and $\big< S_{\bf i}^z\big>_0$ is indicated
for each case (assuming e.g. $\sigma_0=\sigma=\uparrow$ as in Fig. 1).
The same quantities in the pure case are shown in Fig. 2f as a reference.
If we assume that the "down spin" impurity lies, let say, on the A
sublattice then the total (integrated) density for the up spin
is significantly larger on the B sublattice according to (\ref{sumrule})
and reflects the local AF spin environment. Specially interesting
new features also appear on the B sites, namely sharp resonances reflecting the
presence of bound states. Indeed, a comparison of Figs. 2a and 1d with Fig. 2f
reveals that some peaks lie below the bottom of the quasi-particle band
of the pure system and are somehow disconnected from the band (at higher
energy). These features do not appear on the A sites since the density
for the up spins is low on these sites. However, we note that these
bound states are rather extented in space and
clearly Fig. 2d shows that their wavefunctions are not just restricted to the
nearest neighbor sites.

We have carefully studied the spin and the spatial symmetries of
these bound states.
Actually, the local density of states is obtained by decomposing the
local hole operator into its symmetric components,
$c_{\bf i,\sigma}=\sum_\alpha ({\cal N}_{\bf i}^\alpha)^{-1/2}
c_{\bf i,\sigma}^\alpha$ where $\alpha$ labels the
irreducible point group representations \cite{note2}
and ${\cal N}_{\bf i}^\alpha$ are
normalization factors such that $\sum_\alpha ({\cal N}_{\bf i}^\alpha)^{-1}=1$.
The calculation of the density of states
is then made separately in each symmetry sector and the various components
added
afterwards with the appropriate weights,
$N_{\bf ii}^{\sigma,\sigma_0}=\sum_\alpha ({\cal N}_{\bf i}^\alpha)^{-1}
N_{\bf ii,\alpha}^{\sigma,\sigma_0}$, where
$N_{\bf ii,\alpha}^{\sigma,\sigma_0}$ is the density of states in the
$\alpha$-symmetry channel defined by substituting the new operators
$c_{\bf i,\sigma}^\alpha$ in (\ref{density}). Note that each of
these symmetry components
satisfies independently the sum rule (\ref{sumrule}).

The low energy peaks observed in Fig. 2 correspond to bound states
of different spatial symmetries as we shall discuss here.
However, as mentioned above, they all appear in the singlet sector
i.e. for $\sigma=\sigma_0$.
For a more quantitative analysis we define the binding energy by,
$
\Delta_B=(E_{1h,0}^{imp}-E_0^{imp})-(E_{1h,0}^{pure}-E_0^{pure}),
\label{boundstate}
$
where the subsript "1h" refers to the single hole GS (with or without the
impurity). $\Delta_B$ corresponds to
the difference between the energy of an impurity and a mobile hole
confined in the same cluster and their energy when they are separated in
two differentt clusters.
The GS energies $E_{1h,0}^{pure}-E_0^{pure}$ of a single hole moving
in a pure AF background have been calculated elsewhere \cite{DP1holeGS} for
the same clusters.
As seen in Fig. 2 the various resonances
are located
within $\Delta_B$ from the bottom of the pure quasiparticle band.
Since $\Delta_B< 0$ these peaks emerge {\it below} the continuum, in the gap
and hence correspond to actual bound states. In other words, an added hole
can gain energy by binding to the impurity in order to reduce the
magnetic energy loss.

Since the phenomena of binding is a fine balance between delocalization
energy and magnetic energy there is naturally a critical value $J_c$ of J
above which it sets in.
In Fig 1a we show $\Delta_B$ vs J for various cluster sizes for the d-wave
channel. The values of $J_c$ for our clusters are
quite small, between $0.1$ and $0.2$.
However, $J_c$ slighly increases (almost uniformly) with system size and
we expect the actual value to be of the order of 0.3.
It is interesting to notice that the binding energy of a single hole to the
impurity is significantly smaller than the binding energy of a moving pair of
holes \cite{DP2holesbinding} as seen in Fig. 3a. Hence, a collective
delocalization of both objects can strengthen even more the attractive
effective
potential.

For increasing coupling J,
d, s and p-wave bound states successively appear \cite{note3}
(negative value of $\Delta_B$) in the range $0.15<J<0.25$
as seen in Fig. 3b.
The largest (absolute) value of the binding energy is always obtained
in the d-wave channel. However, note that above a rather {\it small}
critical value of $0.25$ (for 20 sites) all bound states of
d, s and p symmetry coexist. As mentioned previously, we expect the
actual critical value of J for the onset of binding to be slightly larger
in the thermodynamic limit.

The distribution of the hole charge density around the impurity
gives useful insights about the bound state wave function.
The charge density on some non-equivalent sites is shown in Fig. 4
for the lowest energy d-wave bound state as a function of J.
Above $\sim J_c$, when binding sets in, the charge density becomes
maximum on the nearest neighbor sites of the impurity.
Hence, the hole wavefunction becomes localized around the impurity
site. We also note that a significant hole density is present on the
4 sites at distance $\sqrt 2$ although this amplitude is not as big as
in the case of a moving hole pair \cite{DP2holespotential}.
A priori, this could seem suprising since the density of states in Fig. 2b
does not show any significant weight at the d-wave bound state energy.
In fact, this simply means that adding the extra hole to the
impurity GS on these particular sites produces a very small
overlap with the actual d-wave bound state. Such a large amplitude on the
{\it next} nearest neighbor sites can only be obtained by adding the hole
at distance 1 from the impurity and by having the system relax to the
its GS configuration. Such retardation effects were also observed in the case
of a bound pair of holes propagating on the lattice \cite{DP2holespotential}.

We conclude by summarizing our findings. In the presence of an inert site, the
extra spin 1/2 created by removing a down spin at site O is distributed.
As seen from the values of $\big< S_{\bf i}^Z\big>_0$ listed in Fig.2 for
$J=0.5$, the total lattice spin deviation of 1/2 spreads largely over the
sites of the $\sqrt{20}\times \sqrt{20}$ cluster. When a hole is added to
the system containing an impurity, d-, s- and p-wave boundstates successively
appear as $J$ increases. The largest binding energy is obtained in the d-wave
channel. The dynamic nature  of this strongly interacting system provides a
collective potential which depends not just on the impurity potential
and the bandstructure hopping parameter t, but rather in an essential way on
the
exchange correlations determined by J. This suggests that the postulated
resonant scattering from $Zn$ impurities introduced in the $CuO_2$ plane of the
cuprates may arise in a natural way from the strong correlations of the host.

\bigskip
\centerline{\bf Acknowledgment}
\bigskip
DJS acknowledges support from the National Science Foundation under grant
DMR92-25027

\newpage
\begin{center}
FIGURE CAPTIONS
\end{center}
\bigskip

\begin{description}

\item{\bf FIG. 1}

Schematic picture of the A-B lattice around the impurity.
A down spin removed at the impurity
site leads to an excess 1/2 spin.
\bigskip

\item{\bf FIG. 2}

Local density of state on various lattice sites at distances $R_{\bf i}$ from
the impurity site on a $\sqrt{20}\times \sqrt{20}$
cluster for J=0.5. The energies of the d, s and p-wave bound states are
indicated by thin dashed lines. The lower edge of the (upper) Hubbard band
of the pure system is shown as a reference by a thicker dashed line.
The expectation values of $S_{\bf i}^z$ on the sites are indicated on the
plots.
\bigskip

\item{\bf FIG. 3}

(a) Binding energy of the d-wave bound state for several cluster
sizes vs J.
The dashed curve corresponds to the binding energy of a pair of
mobile holes on a 26-site cluster in an AF background
(see Ref. \cite{DP2holesbinding}).
(b) Binding energy of the d, s and p-wave bound states on a 20-site cluster
vs J.  The dashed curve corresponds to the binding energy of a pair of
mobile holes on the same cluster (see Ref. \cite{DP2holesbinding}).
\bigskip

\item{\bf FIG. 4}

Hole density on the different non-equivalent 20-site cluster sites
vs J. An extra hole has been introduced into the cluster
in addition to the impurity.
\bigskip

\end{description}


\begin{references}
\bibitem{note2} Note that the number of irreducible representations
depend on the site location and the cluster shape.
\bibitem{note3} Although, reflection symmetries are irrelevant on
most clusters, the $4\times 4$ data suggest that
the d- and s-wave wavefunctions have $d_{x^2-y^2}$ and
ordinary s symmetries.
\end{references}
\end{document}